\documentclass[
 preprint,
superscriptaddress,
 amsmath,
 amssymb,
 aps,
 prl,
]{revtex4-2}

\usepackage{graphicx}
\usepackage[T1]{fontenc}
\usepackage{dcolumn}
\usepackage{bm}
\usepackage[colorlinks,citecolor=red,urlcolor=blue,bookmarks=false,hypertexnames=true,dvipdfm]{hyperref}

\begin{document}

\preprint{YSR2020}

\title{Spin-polarized Yu-Shiba-Rusinov states in an iron based superconductor}
\author{Dongfei Wang}
\email{dwang@physnet.uni-hamburg.de}
\affiliation{Department of Physics, University of Hamburg, Jungiusstrasse 11, 20355, Hamburg, Germany}
\author{Jens Wiebe}
\affiliation{Department of Physics, University of Hamburg, Jungiusstrasse 11, 20355, Hamburg, Germany}
\author{Ruidan Zhong}
\altaffiliation[Present address: ]{Tsung-Dao Lee Institute \& School of Physics and Astronomy, Shanghai Jiao Tong University, Shanghai 200240, China}
\affiliation{Condensed Matter Physics and Materials Science Department, Brookhaven National Laboratory,\\
  Upton, NY 11973, USA}
\author{Genda Gu}
\affiliation{Condensed Matter Physics and Materials Science Department, Brookhaven National Laboratory,\\
  Upton, NY 11973, USA}
\author{Roland Wiesendanger}%
\email{wiesendanger@physnet.uni-hamburg.de}
\affiliation{Department of Physics, University of Hamburg, Jungiusstrasse 11, 20355, Hamburg, Germany}

\date{\today}

\begin{abstract}
Yu-Shiba-Rusinov (YSR) bound states appear when a magnetic atom interacts with a superconductor. Here, we report on spin-resolved spectroscopic studies of YSR states related with Fe atoms deposited on the surface of the topological superconductor FeTe$_{0.55}$Se$_{0.45}$ using a spin-polarized scanning tunneling microscope. We clearly identify the spin signature of pairs of YSR bound states at finite energies within the superconducting gap having opposite spin polarization as theoretically predicted. In addition, we also observe zero-energy bound states for some of the adsorbed Fe atoms. In this case, a spin signature is found to be absent indicating the absence of Majorana bound states associated with Fe adatoms on FeTe$_{0.55}$Se$_{0.45}$.
\begin{description}
\item[Keywords]
Yu-Shiba-Rusinov states, Majorana bound states, Fe-based superconductor, spin-polarized STM
\end{description}
\end{abstract}

\maketitle



When a magnetic atom is interacting with a superconductor having an energy gap $\Delta$ and the temperature is sufficiently low($k_{B}T<\Delta$), Yu-Shiba-Rusinov state(YSRs)  will appear inside the superconducting gap \cite{yu1965bound,shiba1968classical,rusinov1969theory,heinrich2018single}. Since the ﬁrst observation of YSR states by low-temperature scanning tunneling spectroscopy (STS) more than twenty years ago \cite{yazdani1997probing}, numerous studies on various aspects of YSR states have been reported, including improvements of the energy resolution by using superconducting STS probe tips \cite{ji2008high,ruby2015tunneling}, the spatial extension of YSR states \cite{menard2015coherent,yang2020observation,scherubl2020large,kim2020long}, their orbital nature \cite{ruby2016orbital,choi2017mapping}, coupling of the impurity spin to the superconductor substrate \cite{franke2011competition,farinacci2018tuning,malavolti2018tunable}, coupling between YSR impurities \cite{ruby2018wave,kezilebieke2018coupled}, and the formation of YSR chains \cite{nadj2014observation,ruby2017exploring,kim2018toward}. However, the spin polarization of YSR states has rarely been explored experimentally \cite{cornils2017spin,jeon2017distinguishing,ruby2017exploring}. On the other hand, the investigation of the spin nature of impurity bound states has become increasingly important because it allows distinguishing topologically non-trivial Majorana bound states from trivial YSR states which accidentally appear very close to zero energy \cite{li2018majorana,jeon2017distinguishing}. Indeed, such very low-energy YSR states were shown to exist for individual Fe adatoms adsorbed on the hcp-sites of a superconducting Re(0001) substrate where the YSR bound state energy was determined to be on the order of tens of $\mu$eV only \cite{kim2018toward,schneider2019magnetism}.

For magnetic atoms interacting with Fe-based superconductors, such as FeTe$_{0.55}$Se$_{0.45}$, it has been theoretically predicted \cite{jiang2019quantum} and experimentally observed \cite{yin2015observation,wang2018evidence} that zero-energy states exist which might be related with Majorana bound states. However, the spin nature of the bound states induced by magnetic atoms on Fe-based superconductors has not been investigated so far.

Here, by using low-temperature spin-polarized scanning tunneling microscopy (SP-STM) \cite{wiesendanger2009spin} and STS, we successfully observe  finite-energy YSR bound states of Fe atoms deposited on a FeTe$_{0.55}$Se$_{0.45}$ surface and demonstrate that these states are spin-polarized as predicted by theory \cite{flatte1997prb,flatte1997prl}. SP-STS measurements of zero-energy bound states, which are found to coexist in this sample system, reveal the absence of spin polarization, in agreement with the existence of a pair of YSR states being very close to the Fermi level, but in contrast to the interpretation as Majorana bound states.



\begin{figure}[!htbp]
   \centering
   \includegraphics[width=1\textwidth]{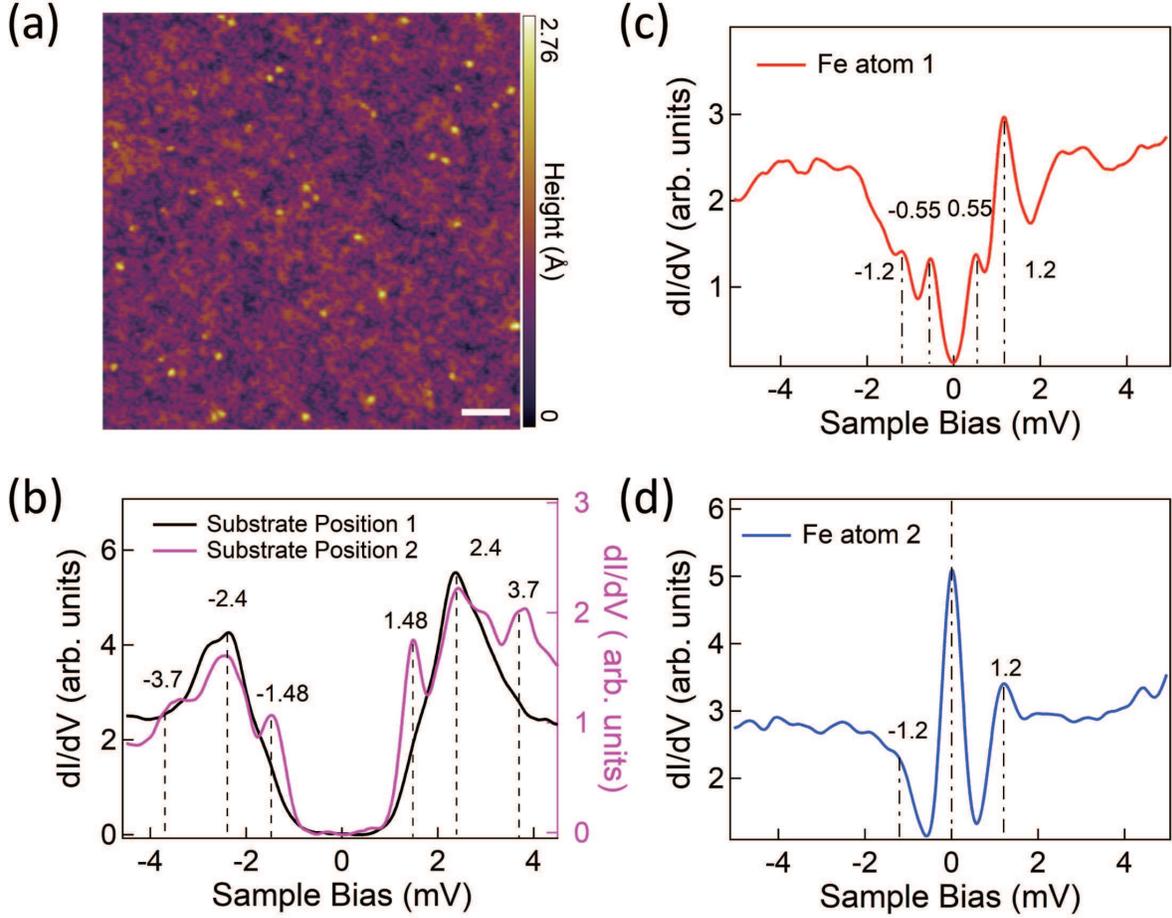}
   \caption{(a) STM topography image of the cleaved FeTe$_{0.55}$Se$_{0.45}$ surface with deposited Fe atoms(yellow spots). Scale bar: 5 nm. (b) Two STS curves taken on clean FeTe$_{0.55}$Se$_{0.45}$ surface regions. (c-d) STS data taken above two different Fe adatoms exhibiting either YSR states at finite energies (c) or a zero-bias peak (d). Tunneling parameters: (a) V=-10 mV, I=200 pA; (b) V$_{stab}$=5 mV, I$_{stab}$=100 pA(black spectrum), V$_{stab}$=-10 mV I$_{stab}$=200 pA(pink spectrum); (c) V$_{stab}$=-10 mV, I$_{stab}$=400 pA; (d) V$_{stab}$=-10 mV, I$_{stab}$=600 pA.}
   \label{fig:Figure1}
\end{figure}

All STM/STS measurements were performed in ultrahigh vacuum at a temperature of 1.1 K and in magnetic ﬁelds up to 3 T applied perpendicular to the sample surface. Differential tunneling conductance(dI/dV) spectra were recorded using a lock-in technique with a bias modulation of 0.03 mV and a frequency of 893 Hz. Before switching off the feedback to record the spectra, the tip is stabilized at a current(I$_{stab}$) and a sample bias voltage(V$_{bias}$) using a tunneling conductance on the order of $10^{-4}2e^{2}/h$. Bulk Cr tips were used for SP-STM measurements. To enhance the spin contrast, sometimes Fe atoms were picked up by the STM tip. Details about tip and sample preparation can be found in ref. \cite{SI}. Individual Fe atoms were deposited in-situ on clean surfaces of freshly cleaved FeTe$_{0.55}$Se$_{0.45}$ single-crystal substrates with a coverage of less than one percent as can be seen in \autoref{fig:Figure1}(a). The adsorbed Fe atoms appear with a mean apparent height of 120 pm in constant-current STM images (see Supplemental Material, Figure S1(b)). Two representative dI/dV spectra, measured on surface regions without Fe adatoms, are shown in \autoref{fig:Figure1}(b), reveal superconducting multi-gap features characteristic for the iron based superconductor FeTe$_{0.55}$Se$_{0.45}$ \cite{miao2012isotropic}. The coherence peaks at $\pm$2.4 mV can be identified clearly in both tunneling spectra, while the coherence peaks at $\pm$1.48 mV only appear as shoulders in the black spectrum. The two sets of coherence peaks reflect the electron pairing within the $\alpha^{'}$ and $\beta$ bands at the $\Gamma$ point. Another pair of peaks at $\pm$3.7 mV appears in some tunneling spectra (see pink curve in \autoref{fig:Figure1}(b)) which may originate from the pairing within the $\gamma$ band. Our observations on the clean FeTe$_{0.55}$Se$_{0.45}$ surface are quite consistent with recent reports on FeTe$_{0.55}$Se$_{0.45}$ \cite{wang2020evidence} and reflects its s-wave like, multi-band pairing nature.

In contrast, the dI/dV spectra measured on top of the deposited Fe atoms on FeTe$_{0.55}$Se$_{0.45}$ show clear in-gap states. Two different types of characteristic tunneling spectra are presented in \autoref{fig:Figure1}(c-d). In \autoref{fig:Figure1}(c) we can clearly identify a pair of in-gap bound states at $\pm$0.55 mV with particle-hole symmetry both with respect to the energy positions as well as the peak heights reminiscent of YSR states. Besides, we can also see sharp peaks at $\pm$1.2 mV with particle-hole symmetry but strong peak height asymmetry. These peaks result from a reduction of the superconducting gap due to the eﬀective magnetic ﬁeld induced by the magnetic Fe adatoms as discussed in the following section, very similar to the reduced gap in the center of superconducting vortices in FeTe$_{0.55}$Se$_{0.45}$ \cite{wang2018evidence}. We also find Fe adatoms exhibiting a zero-bias peak (ZBP) with a full width at half maximum (FWHM) of 0.63 mV as shown in \autoref{fig:Figure1}(d). The observed FWHM reflects the energy resolution of our STM instrument at the measurement temperature of 1.1 K, rather than an intrinsic peak width, and corresponds to the total superconducting gap edge broadening in the tunneling spectra of \autoref{fig:Figure1}(b). Such ZBPs in tunneling spectra have also previously been reported for Fe atoms interacting with superconducting FeTe$_{0.55}$Se$_{0.45}$ and assigned to Majorana bound states associated with the topologically non-trivial surface states of FeTe$_{0.55}$Se$_{0.45}$ \cite{jiang2019quantum,yin2015observation}. However, the spin-polarization of such ZBPs, which can lead to more insight into the nature of such states, has not been explored so far.

\begin{figure}[!htbp]
   \centering
   \includegraphics[width=1\textwidth]{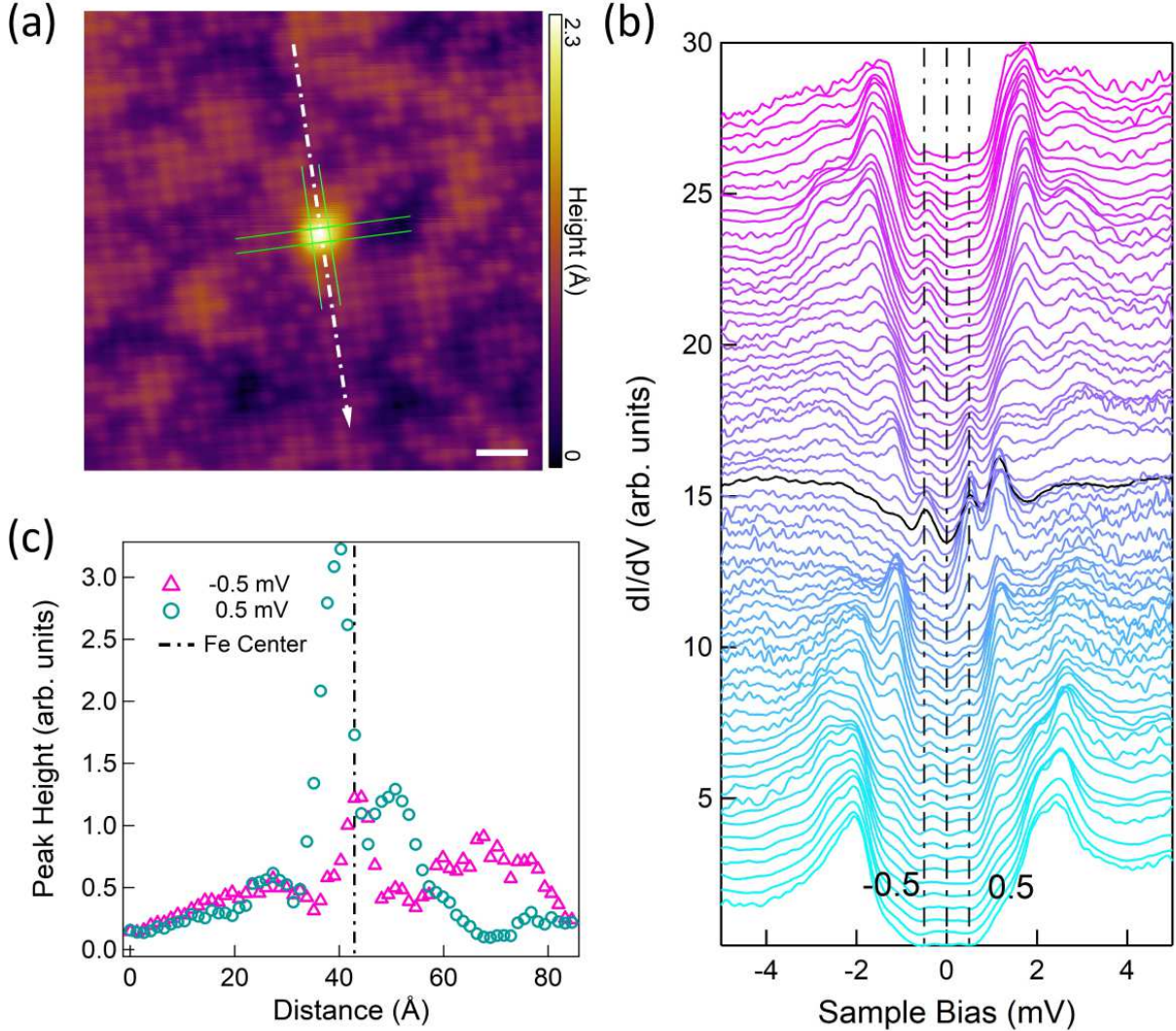}
   \caption{(a) Atomic-resolution STM topography image of the FeTe$_{0.55}$Se$_{0.45}$  surface including an Fe adatom (bright spot) exhibiting finite-energy YSR states. The white dashed arrow indicates the line with a length of 8.45 nm along which dI/dV spectra in (b) have been taken and the green lines are on top of the Se/Te atoms in two different directions. Scale bar: 1 nm. (b) From the bottom to top, a total of 66 dI/dV spectra are presented taken along the line shown in Fig. 2(a). The black spectrum has been recorded above the center of the Fe atom. The black dashed vertical lines at $\pm$0.5 meV indicate the energy position of the YSR$+$ and YSR$-$ peaks. (c) The peak amplitudes of YSR$+$(circle) and YSR$-$(triangle) states extracted from the dI/dV spectra shown in Fig. 2(b). The black dashed vertical line indicates the center position of the Fe atom. Tunneling parameters in (a-b): V$_{stab}$=-10 mV, I$_{stab}$=400 pA.}
   \label{fig:Figure2}
\end{figure}

In the following, we first focus on the spatial distribution of the YSR states which we investigated by taking dI/dV spectra along a line-cut through an Fe atom as illustrated in \autoref{fig:Figure2}(a). The corresponding results are displayed in \autoref{fig:Figure2}(b). From the STM topography image of \autoref{fig:Figure2}(a) we can see that the Fe atom is located at the center position of the Se/Te square lattice as indicated by the solid green lines. Right above the Fe atom, the dI/dV spectrum(black curve in \autoref{fig:Figure2}(b)) shows two YSR states(YSR$+$ and YSR$-$) with particle-hole symmetry. As our tunnel junction conductance is on the order of $10^{-4}2e^{2}/h$, Andreev reflection processes \cite{ruby2015tunneling} can be ruled out. Hence the in-gap states we observe here should reflect the intrinsic scattering of Cooper pairs by the Fe atom. We also notice a considerable asymmetry regarding the peak heights of the YSR states. Due to the weak tunneling conditions in our experiment, this asymmetry can either be attributed to the particle-hole asymmetry of the bands near the Fermi energy or to the spin-independent Coulomb potential scattering processes \cite{flatte1997prb,balatsky2006impurity}. We can see this asymmetry more clearly in \autoref{fig:Figure2}(c) where the observed peak height values of the YSR$+$ and YSR$-$ states as a function of distance from the Fe atom are extracted.

Another important characteristic feature of YSR states is their spatial decay behavior. From \autoref{fig:Figure2}(c) we can see that both bound states at $\pm$0.5 meV exhibit a decay length of about 4 nm. If we use the model for 3D case \cite{balatsky2006impurity}, the decay length $r_{3d} \sim \xi_{0}/\sqrt{1-\epsilon^{2}}$ where $\xi_{0}$ is the coherence length and $\epsilon$ is defined as ratio between the YSRs state energy $E_{\text{YSR}}$ and the superconducting gap $\Delta$. In our case, $r_{3d}$ results in a value of 2.08 nm, if we take $\xi_{0}=2$ nm \cite{jiang2019quantum}, $E_{\text{YSR}}=0.5$ meV and $\Delta=1.8$ meV. This theoretical value is smaller than the experimentally observed one. A possible reason could be that the Fe impurity primarily interacts with the surface state of FeTe$_{0.55}$Se$_{0.45}$ which would mean that we have to consider the 2D case. While the YSR states exhibit a $1/r^{2}$ decay for the 3D case, a 1/r decay is expected for the 2D case \cite{menard2015coherent}. Therefore, considering an effective 2D system, we can expect a decay length $r_{2d}=r_{3d}^2=4.3$ nm, which is in good agreement with our experimental value of 4 nm. In contrast, the decay length of the peaks at $\pm$1.2 mV is only about 0.5 nm which is much smaller than for the YSR state at $\pm$0.5 mV. This fast decay behavior can be attributed to the influence of a magnetic dipole field induced by the Fe atom. A previous study showed that an Fe atom can have a dipole moment of $m=5.4\mu_{B}$ when placed on the MgO surface exhibiting a $1/r^{3}$ decay behavior \cite{choi2017atomic}. If we substitute such value into the dipole field equation $B(r)=\mu_{0}m/4\pi r^{3}$, we will get a field strength of 320 mT at 0.25 nm and 40 mT at 0.5nm. This is comparable to our experimentally observed decay length of 0.5 nm beyond which the dipole magnetic field will not strongly affect superconductivity.

\begin{figure}[!htbp]
   \centering
   \includegraphics[width=1\textwidth]{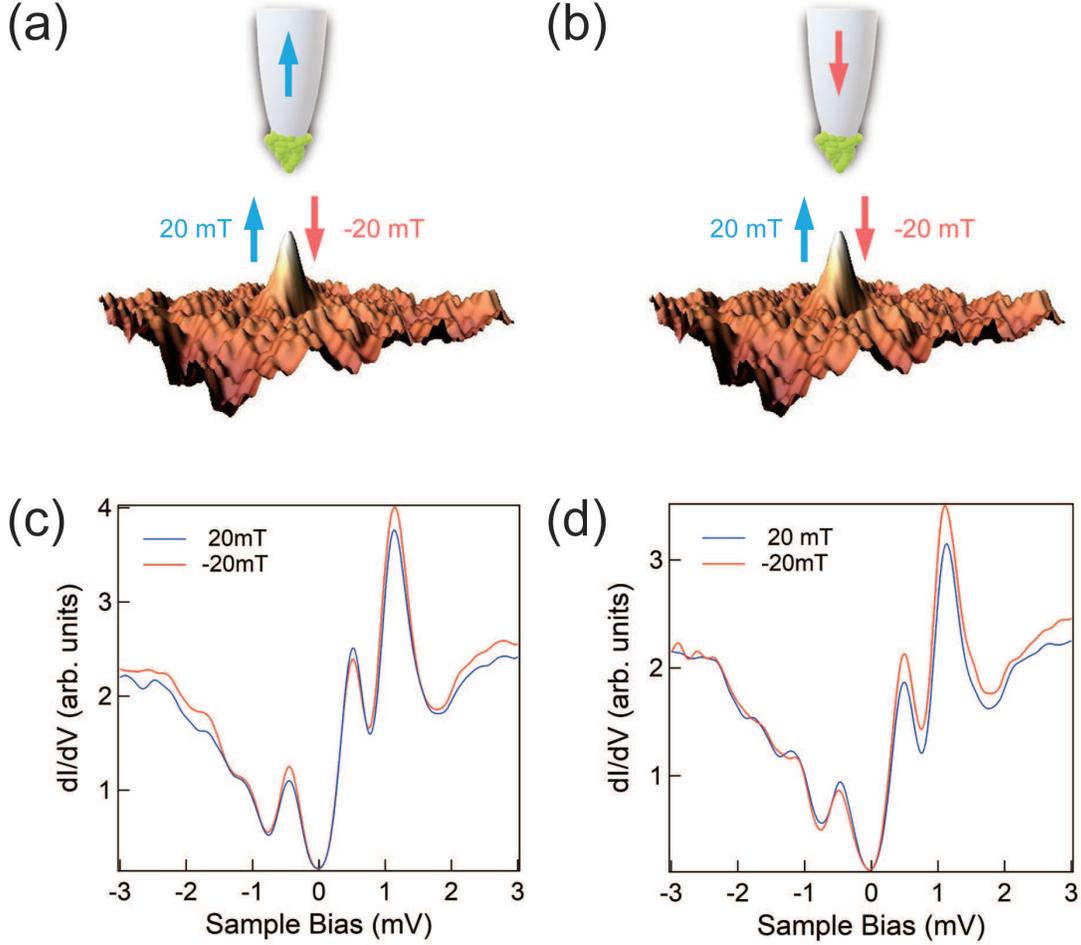}
   \caption{(a-b) Illustration of the spin-polarized tunneling experiment on individual Fe adatoms on a FeTe$_{0.55}$Se$_{0.45}$ surface. The SP-STM tip is magnetized up in (a) and down in (b) as indicated by the blue and red arrows, respectively. The magnetic moment orientation of the Fe atom is controlled by a very small external magnetic ﬁeld of $\pm$20 mT. The tip magnetization orientation remains unchanged in such a small external ﬁeld. (c-d) The spin-resolved dI/dV spectra taken at an external field of $\pm$20 mT, when the tip magnetization has been prepared up (c) or down (d). All spectra were recorded above the center of the Fe atom. Tunneling parameters in (a-b): V$_{stab}$=-10 mV, I$_{stab}$=400 pA.}
   \label{fig:Figure3}
\end{figure}

After making sure that the observed in-gap states at $\pm0.5$ mV are YSR bound states, we performed spin-resolved tunneling spectroscopy measurements in order to probe their spin-dependent properties. As illustrated in \autoref{fig:Figure3}(a-b), the magnetization direction of an Fe-decorated Cr tip is controlled by applying a sufficiently large external magnetic ﬁeld of $\pm$1.5 T following the procedure as described in ref. \cite{cornils2017spin}. Details about the characterization of the spin sensitivity of the SP-STM probe tip can be found in the Supplemental Material(Figure S2). After preparing the SP-STM tip in a well defined magnetization state, we polarized the Fe atom in two opposite directions by applying a small external magnetic ﬁeld of $\pm$20 mT. Such a small magnetic filed does not affect the magnetization state of the SP-STM tip as demonstrated in ref. \cite{cornils2017spin}. The results of the spin polarized tunneling experiments are displayed in \autoref{fig:Figure3}(c) and (d). We first focus on the magnetic field response of the YSR$+$ peak at $+$0.5 mV. In the case of the upward magnetized probe tip(previously polarized at $+$1.5 T), the dI/dV signal above the Fe atom polarized at $+$20 mT is higher than that for the Fe atom being polarized at $-$20 mT as shown in \autoref{fig:Figure3}(c). However, the behavior of the dI/dV signal reverses when the tip has been downward magnetized at $-$1.5 T(see \autoref{fig:Figure3}(d)). The fact that a different magnitude of the dI/dV signal is always obtained when tip and sample are polarized in the same direction as compared to the case where they are polarized in opposite direction provides clear evidence for the observation of spin-polarized tunneling. On the other hand, according to theory \cite{flatte1997prl}, the sign of the spin polarization of the YSR$-$ state should be opposite to the one of the YSR$+$ state. Therefore, we can expect that the dI/dV signal of the YSR$-$ state exhibits an inverse response to a change of the magnetization direction of the SP-STM probe tip compared to that of the YSR$+$ state. This is indeed observed in \autoref{fig:Figure3}(c), where the YSR$-$ peak is higher for the Fe atom being polarized in a field of $-$20 mT, while the YSR$+$ peak is higher for an applied field of $+$20 mT. Similar behaviour is also found for the data displayed in \autoref{fig:Figure3}(d).

\begin{figure}[!htbp]
   \centering
   \includegraphics[width=0.9\textwidth]{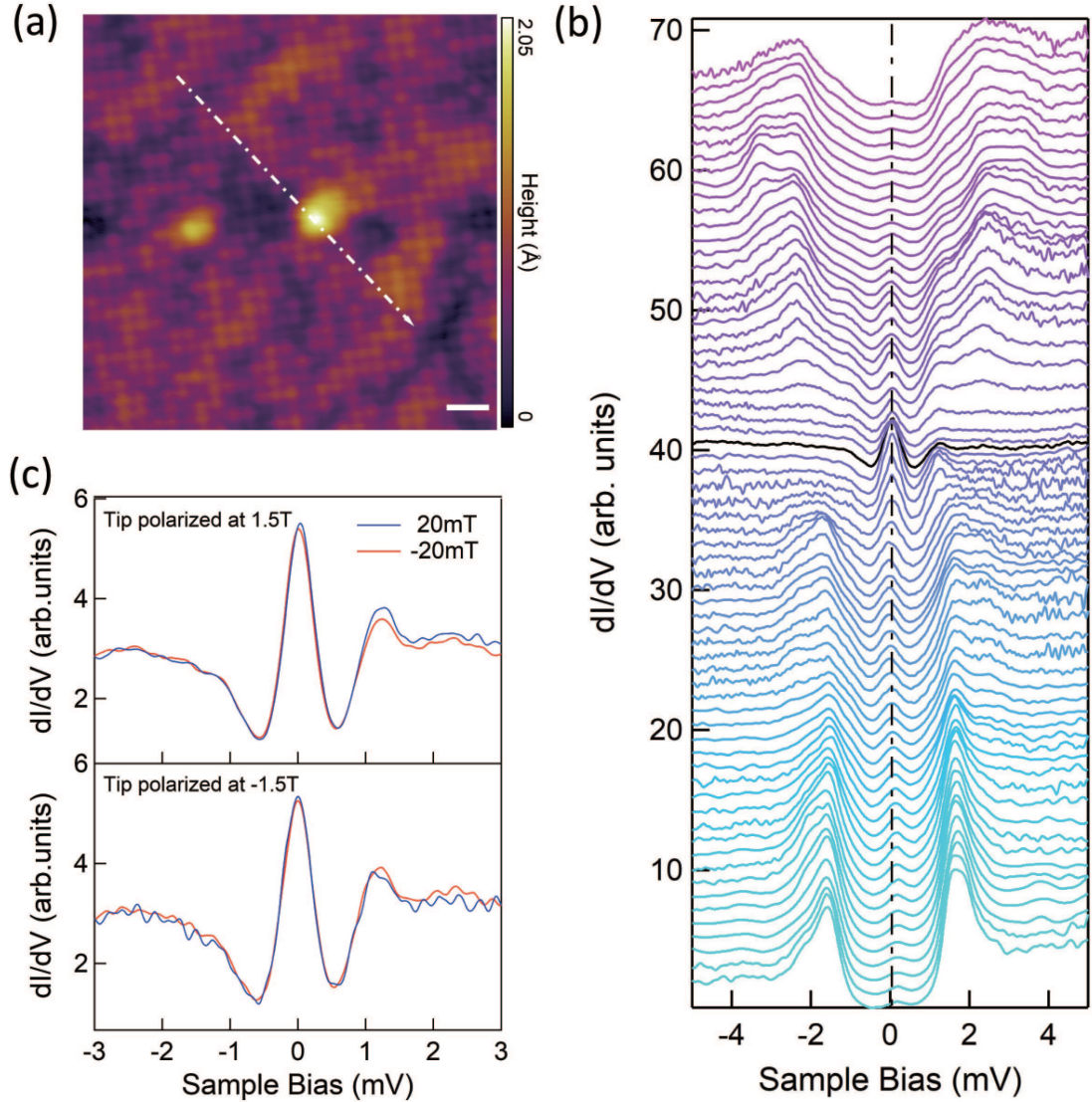}
   \caption{(a) STM topography image of the FeTe$_{0.55}$Se$_{0.45}$ surface including an Fe atom exhibiting a zero-bias peak (ZBP). The white dashed arrow indicates the line with a length of 8.34 nm along which the dI/dV spectra in (b) have been taken. Scale bar: 1 nm. (b) From the bottom to top, a total of 66 dI/dV spectra are displayed taken along the line shown in Fig. 4(a). The black spectrum was measured above the center of the Fe atom. (c) Magnetic ﬁeld dependent measurements of the dI/dV spectrum exhibiting a ZBP. The SP-STM tip has been polarized up or down in an external magnetic field of $\pm$1.5 T. The orientation of the Fe atom’s magnetic moment is controlled by a small external magnetic ﬁeld of $\pm$20 mT which does not affect the magnetization direction of the SP-STM probe tip. All tunneling spectra were measured above the center of the Fe atom. Tunneling parameters in (a-b): V$_{stab}$=-10 mV, I$_{stab}$=600 pA.}
   \label{fig:Figure4}
\end{figure}

Finally, by using the same spin-polarized probe tip, we examined the magnetic field response of the ZBP which is frequently observed for individual Fe atoms interacting with the FeTe$_{0.55}$Se$_{0.45}$ surface. \autoref{fig:Figure4}(a) shows an STM topography image including an Fe adatom exhibiting such a ZBP. Tunneling spectroscopic data has been obtained along the line indicated in that figure. The corresponding results are displayed in \autoref{fig:Figure4}(b). From the set of 66 dI/dV spectra we can see that this ZBP does not shift in energy over quite a very large distance from the Fe atom. We then performed spin-resolved tunneling spectroscopy measurements in a similar way as described before for the finite-energy YSR bound states at $\pm$0.5 mV. From the experimental data shown in \autoref{fig:Figure4}(c) we can see that within the experimental error, the dI/dV signal at zero bias does not respond to a change of polarization of the Fe atom nor to a change of the tip magnetization state. This obvious absence of a measurable spin polarization can be understood if the ZBP would result from a pair of YSR states being very close in energy and very near the Fermi level such that spin-up quasiparticles mix with spin-down quasiparticles. Recently, ZBPs observed for such Fe atoms interacting with FeTe$_{0.55}$Se$_{0.45}$ have been interpreted as Majorana bound states(MBS) induced by quantum anomalous vortices \cite{jiang2019quantum}. In that case, however, spin-selective Andreev reﬂection processes should be observable as in the case of MBS inside real vortices \cite{kawakami2015evolution,sun2016majorana}. However, the absence of a spin signature of the ZBPs as observed in our experiments indicate that they most likely originate from YSR states being accidentally located very close to the Fermi level, such as reported before for Fe adatoms on hcp-sites of a Re(0001) surface \cite{kim2018toward,schneider2019magnetism}, rather than from a MBS.

In conclusion, using spin-polarized STM/STS, we revealed the spin nature of YSR states of Fe atoms on a superconducting FeTe$_{0.55}$Se$_{0.45}$ substrate. Our experimental results obtained for finite-energy YSR states are consistent with the theoretical prediction of opposite signs of spin polarization for the electron- and hole-like components. We also investigated the spin nature of the ZBP and found no spin-dependent response, in disagreement with interpretations of such ZBP as Majorana bound states.

We would like to thank Thore Posske for useful discussions as well as Torben H\"anke and Anand Kamlapure for technical support. We also thank Hong Ding, Fazhi Yang and Cuihua Liu for providing the samples. This work has been supported by the EU via the ERC Advanced Grant ADMIRE (No. 786020), the DFG via the Cluster of Excellence “Advanced Imaging of Matter” (EXC 2056, project ID 390715994) and the US Department of Energy, oﬃce of Basic Energy Sciences (contract no. de-sc0012704).

\bibliographystyle{apsrev4-2}
\bibliography{ms}

\end{document}


\preprint{YSR2020SI}

\title{Supplemental Material: Spin-polarized Yu-Shiba-Rusinov states in an iron based superconductor}
\author{Dongfei Wang}
\email{dwang@physnet.uni-hamburg.de}
\affiliation{Department of Physics, University of Hamburg, Jungiusstrasse 11, 20355, Hamburg, Germany}
\author{Jens Wiebe}
\affiliation{Department of Physics, University of Hamburg, Jungiusstrasse 11, 20355, Hamburg, Germany}
\author{Ruidan Zhong}
\altaffiliation[Present address: ]{Tsung-Dao Lee Institute \& School of Physics and Astronomy, Shanghai Jiao Tong University, Shanghai 200240, China}
\affiliation{Condensed Matter Physics and Materials Science Department, Brookhaven National Laboratory,\\
  Upton, NY 11973, USA}
\author{Genda Gu}
\affiliation{Condensed Matter Physics and Materials Science Department, Brookhaven National Laboratory,\\
  Upton, NY 11973, USA}
\author{Roland Wiesendanger}%
\email{wiesendanger@physnet.uni-hamburg.de}
\affiliation{Department of Physics, University of Hamburg, Jungiusstrasse 11, 20355, Hamburg, Germany}

\date{\today}

\maketitle


\section{SAMPLE PREPARATION}

\begin{figure}[!htbp]
   \centering
   \includegraphics[width=1\textwidth]{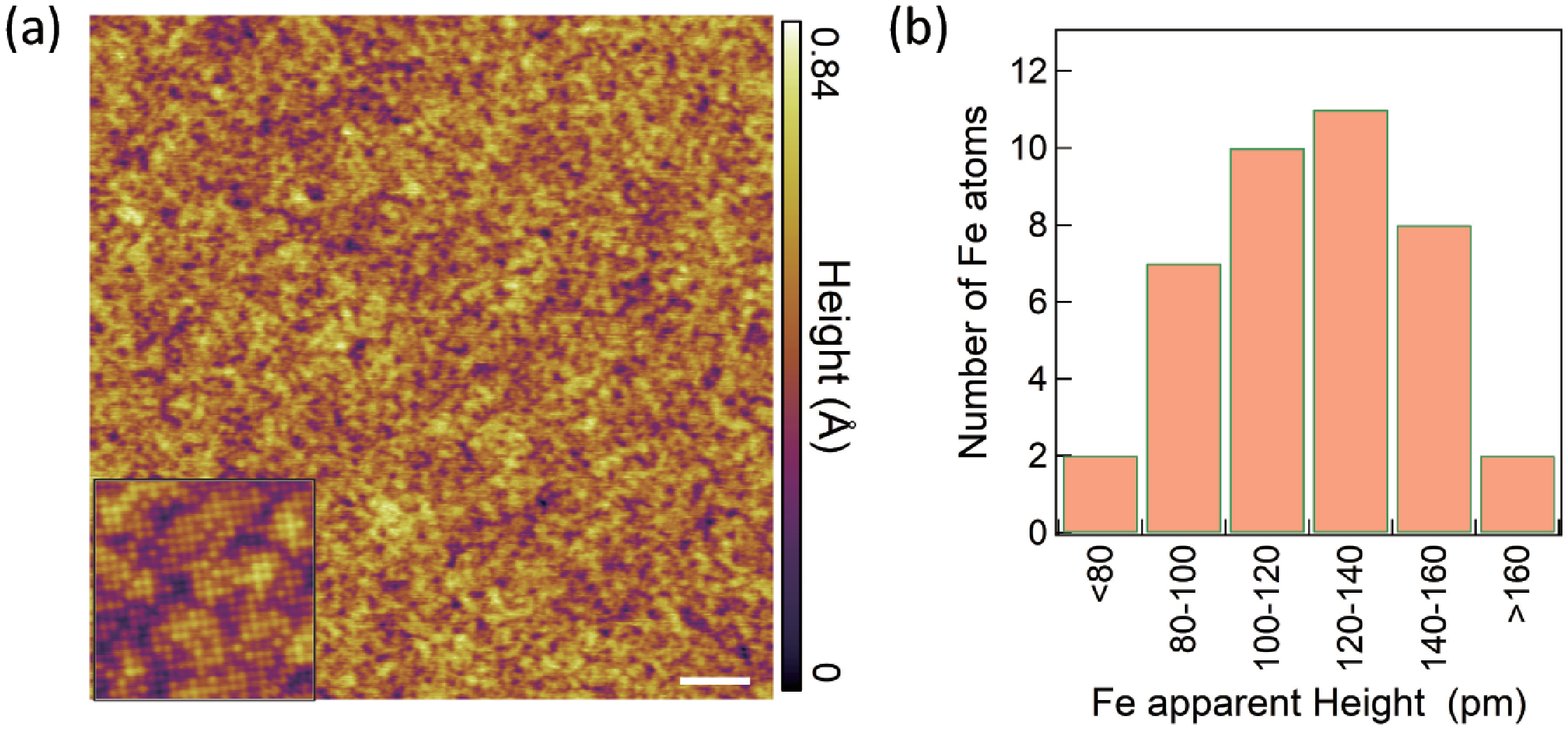}
   \caption{(a) Large-scale STM topography image of the cleaved FeTe$_{0.55}$Se$_{0.45}$ surface. Scale bar: 10 nm. Inset: Zoom-in atomic-resolution STM image (8 nm$\times$8 nm). (b) Statistics of the apparent height of the deposited Fe atoms shown in Figure 1(a). Tunneling parameters in (a): V=-1 V, I=100 pA. Inset: V=-10 mV, I=100 pA.}
   \label{fig:FigureS1}
\end{figure}

High-quality  FeTe$_{0.55}$Se$_{0.45}$ samples were grown using the self-ﬂux method. Their superconducting transition temperature Tc is 14.5 K \cite{wen2009short}. The samples were cleaved in-situ by vacuum tape in the preparation chamber with a vacuum better than $5\times10^{-10}$ mbar and then transferred to the low-temperature ultra-high vacuum (UHV) STM. Large clean surface areas of more than 100 nm across allowing for atomic-resolution STM imaging can be obtained as shown in \autoref{fig:FigureS1}(a). A low density of Fe atoms was then deposited in-situ by e-beam evaporation onto the clean FeTe$_{0.55}$Se$_{0.45}$ surface held at 4 K. The resulting Fe atom coverage is less than 1 percent. We analyzed the apparent heights of the deposited Fe atoms for the field of view shown in Figure 1(a), leading to the statistics presented in \autoref{fig:FigureS1}(b). The apparent heights of the Fe adatoms in the STM images amounts to 80 - 160 pm using the tunneling parameters: V=-10 mV, I=200 pA.

\section{SP-STM TIP CALIBRATION}

\begin{figure}[!htbp]
   \centering
   \includegraphics[width=1\textwidth]{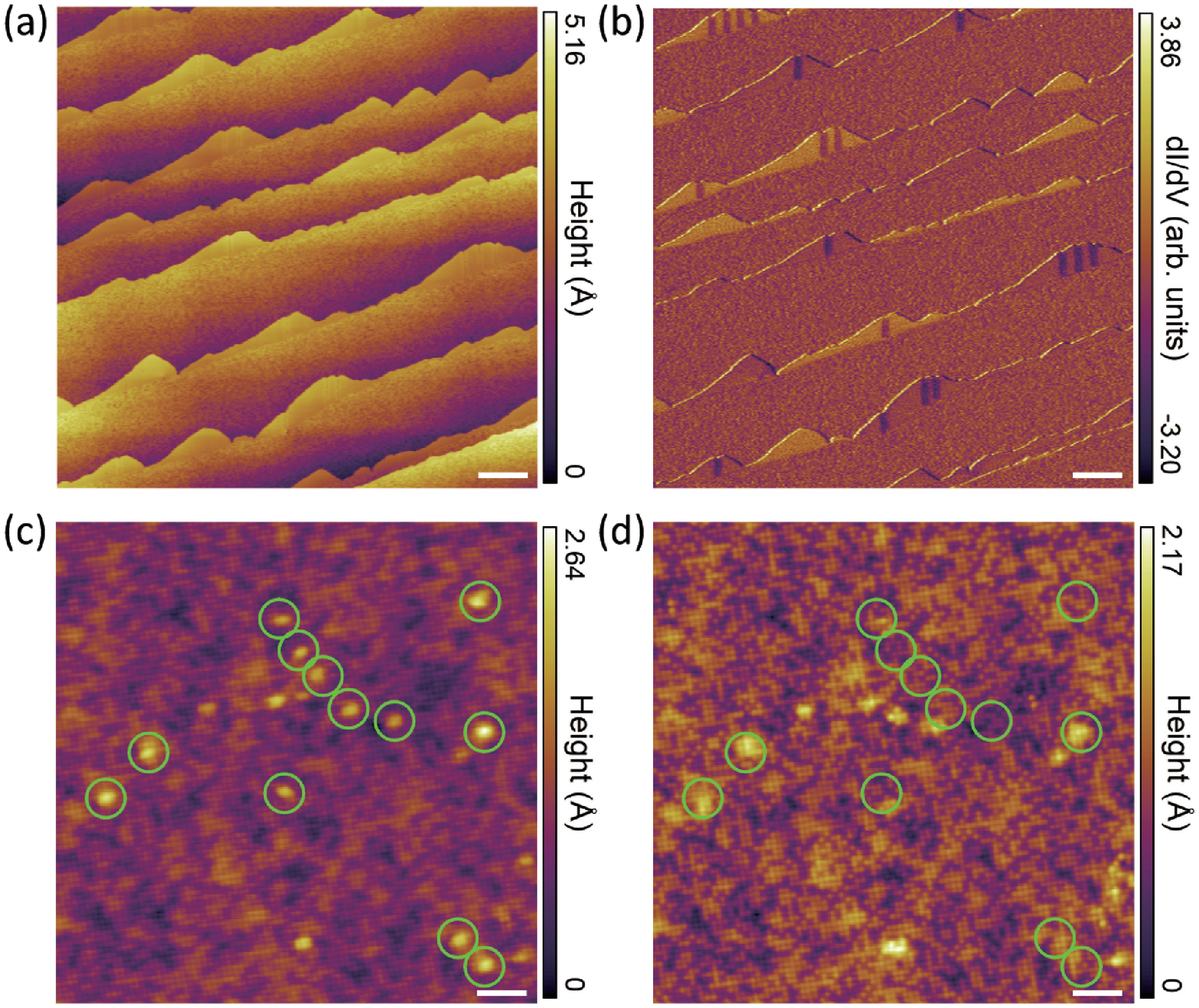}
   \caption{(a) STM topography image of Fe double-layer stripes grown on a stepped W(110) substrate. (b) Corresponding spin-resolved dI/dV image obtained at V$_{stab}$=100mV. The bright and dark regions correspond to magnetic domains of opposite out-of plane magnetization in the Fe double-layer. (c-d) STM topography image of the same area before (c) and after (d) picking up Fe atoms by the Cr tip. Green circles indicate where the vertical atom manipulation took place. Tunneling parameters in (a-b): V=0.1V, I=100pA. (c-d) V=-10 mV, I=1 nA.  The lock-in modulation used in (b) was 13 mV at 4370 Hz. Scale bar for (a-b): 30 nm; for (c-d): 3 nm.}
   \label{fig:FigureS2}
\end{figure}

The tip used for our SP-STM study is a bulk Cr tip prepared by electrochemical etching in a NaOH solution \cite{schlenhoff2010bulk}. After loading into our UHV-system, the Cr tips were flashed above 1500 K to remove surface oxide layers. Then the spin sensitivity of the prepared SP-STM tips was checked by resolving the well known magnetization pattern of double-layer Fe nanostructures on a stepped W(110) substrate \cite{kubetzka2003spin}. A clean W(110) surface was obtained by heat treatment in an oxygen atmosphere at around 1500 K followed by ﬂash-annealing at 2300 K for several times  \cite{bode2007preparation}. Fe was then deposited by e-beam evaporation onto the clean W(110) surface in a vacuum better than $4\times10^{-10}$ mbar. Subsequent annealing near 500 K resulted in a step-flow growth of Fe on the stepped W(110) substrate yielding Fe double-layer stripes close to the W(110) step edges (\autoref{fig:FigureS2}(a)). Previous studies indicated that the Fe double-layer stripes on W(110) have an out-of-plane magnetization with a characteristic domain structure \cite{kubetzka2003spin}. We can observe such spin-dependent contrast by SP-STM imaging at 100 mV using the bulk Cr tip as shown in \autoref{fig:FigureS2}(b). In order to achieve high spin contrast the Cr-tip was occasionally treated by several high-voltage pulses against the Fe/W(110) sample.

After confirming the spin sensitivity of the Cr tip, we immediately changed the sample to FeTe$_{0.55}$Se$_{0.45}$ with adsorbed Fe atoms. Previous studies showed that the tip’s spin polarization can be significantly enhanced by picking up some magnetic atoms to its apex \cite{wiesendanger2009spin}. Here we use the same method as described in ref. \cite{cornils2017spin} to achieve a high spin polarization as well as a high spatial resolution. \autoref{fig:FigureS2}(c) and (d) show the same surface area before and after picking up Fe atoms from the sample, respectively. Obviously, the spatial resolution is highly improved in \autoref{fig:FigureS2}(d). In total, we picked up more than 50 Fe atoms to the Cr tip apex. Usually, with such a tip configuration, the critical field needed to change the direction of tip magnetization is between 0.8 T and 1 T. Therefore, we used a ﬁeld of 1.5 T to make sure that our tip can be remagnetized by the external magnetic ﬁeld.

\bibliographystyle{apsrev4-2}
\bibliography{supplement}